\documentclass[12pt]{article}
\usepackage{geometry}
\usepackage[applemac]{inputenc}
\usepackage{amsmath}
\usepackage{graphicx}
\usepackage[colorlinks]{hyperref}

\title{A Lorentz Invariant Phenomenological Model of Quantum Gravity}
\date{}

\author{Yuri Bonder\footnote{In collaboration with D. Sudarsky and A. Corichi. Author's  e-mail: yuri.bonder@nucleares.unam.mx}\\
Instituto de Ciencias Nucleares \\
Universidad Nacional Aut\'{o}noma de M\'{e}xico\\
A. Postal 70-543, M\'exico D.F. 04510, M\'exico}
 
\begin{document}

\maketitle

\section*{Abstract}

We consider a model of Quantum Gravity phenomenology, based on the idea that space-time may have some unknown granular structure that respects the Lorentz symmetry. The proposal involves non-trivial couplings of curvature to matter fields and leads to a well defined phenomenology. In this manuscript, a brief description of the model is presented together with some results obtained using linearized gravity and the Newtonian limit, which could be useful when comparing with real experiments.

\section{Motivation}

It is clear that current and future particle accelerators will not be reaching the Plank Energy regime anytime soon and thus Quantum Gravity could well be beyond our experimental reach. Recently, people have considered the possibility that Quantum Gravity effects may be detected through violations of the space-time symmetries, particularly Lorentz symmetry. The search for Quantum Gravity phenomenology through violations of Lorentz symmetries can be traced to \cite{Initial-LIV, SME}. However, this possibility faces now very serious experimental bounds \cite{Bounds} and some theoretical challenges. For instance, a preferential frame, which would be naturally associated with space-time discreetness, when combined with the radiative corrections that appear in Quantum Field Theory leads to the prediction of large effects that have not been observed \cite{Collins}.

In this work we deal with the following issue: Assuming that the space-time has some granular structure whose exact form is unknown, what are the other possible ways, besides Lorentz symmetry violations, that Quantum Gravity might become manifest? In order to offer some answer we consider a model based on the assumption that there are space-time building blocks which are Lorentz invariant, ensuring that the theory respects the large scale Lorentz symmetry. Then we consider possible phenomenological signatures which could  arise in such situation. In order to proceed, and given that the fundamental theory describing such structure (Quantum Gravity theory) is unknown, we base our considerations on some symmetry principles; in particular, on an analogy with solid state physics (the detailed proposal can be found in \cite{NewQGP, QGPwithoutLSV}).

The basic idea is the folowing: When the macroscopic symmetry of a crystal coincides with the symmetry of the fundamental crystalline cells, no deviation from the macroscopic symmetry can be detected as indication of the fundamental structure of the crystal. On the other hand, if both symmetries, the macro and the microscopic, do not coincide, deviations from the macroscopic symmetry are expected to be present. For example, in a macroscopic sphere made of a cubic crystal, the mismatch of the macro and the microscopic symmetry would manifest through the roughness at the sphere's surface. 

This idea is translated to space-time by considering that its granular structure plays the role of the crystalline cell and the Lorentz symmetry is analogous to the fundamental crystalline symmetry. Following this analogy, in a flat region of space-time the granular structure is not expected to become manifest (in particular not through the breakdown of Lorentz symmetry). On the other hand, in regions of space-time where the Riemann tensor $R_{\mu\nu\rho\sigma}$ (which measures the failure of an open region of space-time to be Minkowski) is not zero, it could be possible to detect the mismatch of symmetries. Thus we search for nontrivial couplings of the Riemann tensor with matter. We note that, as the Ricci tensor $R_{\mu\nu}$ is locally determined by the energy-momentum tensor of the matter fields, the coupling of $R_{\mu\nu}$ to the matter fields looks like a self-interaction and is therefore uninteresting. We will therefore focus our attention on the traceless 
 part of the Riemann tensor, namely the Weyl tensor $W_{\mu\nu\rho\sigma}$. Thus, we consider nontrivial couplings of the Weyl tensor to matter fields, in particular, to fermions.

\section{The proposal: Coupling Weyl to matter}

We seek for coupling terms of Weyl with fermion fields that are minimally suppressed by Planck's mass $M_{Pl}$. Recall that the $n$-dimensional Lagrangian term (in mass dimensions) must be suppressed by $(1/M_{Pl})^{n-4}$, thus, the dominant part of the coupling is expected to have dimension $5$. It is well known that the Weyl tensor has mass dimension $2$ and the fermionic fields have mass dimension $3/2$. Needless to say that the coupling terms we are looking for must also be scalar under Lorentz transformations.

The most obvious $5$-dimensional coupling term has the form 
\begin{equation} \label{obv coupling}
W_{\mu\nu\rho\sigma}\bar{\psi} \gamma^\mu\gamma^\nu \gamma^\rho\gamma^\sigma\psi,
\end{equation}
where $\psi$ stands for the various fundamental spinor fields in the standard model and $\gamma^\mu$ are the
Dirac matrices. Considering that derivatives have mass dimension $1$ and that Weyl is traceless, is easy to show that all the $5$-dimensional coupling of Weyl with fermions have the form of (\ref{obv coupling}). Given that,
\begin{equation} \label{1er coup}
W_{\mu\nu\rho\sigma}\bar{\psi} \gamma^\mu\gamma^\nu \gamma^\rho\gamma^\sigma\psi=W_{\mu\nu\rho\sigma}\epsilon^{\mu\nu\rho\sigma}\bar{\psi} \gamma^5\psi,
\end{equation}
where $\epsilon^{\mu\nu\rho\sigma}$ is the volume $4$-form, this coupling vanishes since the Weyl tensor have no totally antisymmetric part. 

Instead of giving up at this point we look for alternatives. One possibility is to consider some object derived from Weyl that has a different index strcuture. This can be achieved by considering the Weyl tensor as a self-adjoint map $\cal{S}\rightarrow\cal{S}$, $\cal{S}$ being the $6$-dimensional space of $2$-forms. In addition, the space-time metric endows the six dimensional vector space
$\cal{S}$ with a pseudo-Riemannian metric $G_{\mu\nu\rho\sigma}$. Due to the fact that this map is self-adjoint, it can be diagonalized and its eigenvalues $\lambda^{(s)}$ and eigenforms $\Xi^{(s)}_{\mu\nu}$ can be used to construct the desired terms.
We will use $G_{\mu\nu\rho\sigma}$ to normalize the non-null eigenforms according to
\begin{equation} \label{GXX}
G^{\mu\nu\rho\sigma}\Xi^{(s)}_{\mu\nu} \Xi^{(s)}_{\rho\sigma}= \pm 1.
\end{equation}
Furthermore, from the Weyl tensor symmetries, it can be seen \cite{QGPwithoutLSV} that it satisfies
\begin{equation}
{\epsilon_{\mu\nu}}^{\rho\sigma} {W_{\rho\sigma}}^{\alpha\beta} {\left( \epsilon^{-1}\right)_{\alpha\beta}}^{\gamma\delta} ={W_{\mu\nu}}^{\gamma\delta},
\end{equation}
which implies that, if $\Xi^{(s)}_{\mu\nu}$ is a Weyl eigenform with negative (or positive) norm, then $\widetilde{\Xi}^{(s)}_{\mu\nu}\equiv{\epsilon^{\rho\sigma}}_{\mu\nu}\Xi^{(s)}_{\rho\sigma}$ is also a Weyl eigenform with the same eigenvalue and the opposite norm. This shows that Weyl is always degenerated and the recipe is ill defined. In order to fix this, one needs to discriminate between all the linear combinations of the degenerated eigenforms, which are all clearly eigenforms with the same eigenvalue. It is noteworthy that the same object that leads to the degeneration, namely $\epsilon_{\mu\nu\rho\sigma}$, can be used to discriminate between all the linear combinations of degenerated eigenforms. This is done by imposing to the three eigenforms with negative norm (represented by $\Xi^{(l)}_{\mu\nu}$, $l=1,2,3$) the condition
\begin{equation} \label{eXX}
\epsilon^{\mu\nu\rho\sigma}\Xi^{(l)}_{\mu\nu} \Xi^{(l)}_{\rho\sigma}=0.
\end{equation}
Note that three eigenforms with positive norm $\widetilde{\Xi}^{(l)}_{\mu\nu}$ are obtained from the eigenforms with negative norm $\Xi^{(l)}_{\mu\nu}$ by contracting with $\epsilon_{\mu\nu\rho\sigma}$. Therefore, the Weyl eigenforms with negative norm satisfying the conditions (\ref{GXX}) and (\ref{eXX}) and the corresponding positive normed eigenforms are uniquely defined\footnote{There is still an ambiguity in the signs of the eigenforms but this signs are absorbed by the coupling constants.} and are the objects that are coupled to fermions.

It is worth noting that the space-time volume $4$-form plays an important role in this scheme, and thus, that we are taking the view that the space-time structure may involve, in addition to the metric and a time orientation, a spacial orientation. Therefore, the suggestion is that gravity at a quantum level may involve violations of discrete symmetries such as spatial inversion and time reversal. There is a known pattern in the interactions of the Standard Model that the weaker the interaction, less symmetries it respects and classical gravity breaks this pattern.

In order to determine the form of the coupling terms, note that Weyl eigenvalues have the dimensions of the Weyl tensor and its eigenforms are dimensionless. The most natural way of writing the dominant part of the coupling of Weyl and fermionic matter fields, taking into the account the unavoidable degeneration mentioned above and a possible flavor dependence, is
\begin{equation} \label{L2}
\mathcal{L}_f=\frac{1}{M_{Pl}}\sum_a \sum_{l=1}^3 \lambda^{(l)}\left(\xi_a^{(l)}\Xi^{(l)}_{\mu\nu}+\widetilde{\xi}_a^{(l)}\widetilde{\Xi}^{(l)}_{\mu\nu}\right)\bar{\psi}_a
\gamma^\mu\gamma^\nu \psi_a,
\end{equation}
where the index $a$ denotes flavor, $\xi_a^{(l)}$ and $\widetilde{\xi}_a^{(l)}$ are dimensionless coupling constants, $\Xi^{(l)}_{\mu\nu}$ are the Weyl eigenforms with negative norm that satisfy equations (\ref{GXX}) and (\ref{eXX}) and $\widetilde{\Xi}^{(l)}_{\mu\nu}={\epsilon_{\mu\nu}}^{\rho\sigma}\Xi^{(l)}_{\rho\sigma}$. This terms can be generalized by using the fact that $\sqrt{\lambda^{(l)}}/M_{Pl}$ is dimensionless and introducing parameters $r$ and $\widetilde{r}$, required to be greater than $-1$, and writing
\begin{equation} \label{L3}
\mathcal{L}_f=\sum_a \sum_{l=1}^3\sqrt{\lambda^{(l)}}\left\lbrace \xi^{(l)}_a \left( \frac{\sqrt{\lambda^{(l)}}}{M_{Pl}}\right)^r \Xi^{(l)}_{\mu\nu} +{\widetilde{\xi}}^{(l)}_a \left( \frac{\sqrt{\lambda^{(l)}}}{M_{Pl}}\right)^{\widetilde{r}} {\widetilde{\Xi}}^{(l)}_{\mu\nu} \right\rbrace\bar{\psi}_a
\gamma^\mu\gamma^\nu \psi_a,
\end{equation}
which coincides with (\ref{L2}) when $r=\widetilde{r}=1$. Expression (\ref{L3}) is the one that is used in the rest of the manuscript. 

\section{Phenomenology ready expressions}

In this work we will only consider the expressions heeded to deal with experiments to be carried out on Earth, thus, we are in the Newtonian regime, where the space-time metric is taken as flat metric $\eta_{\mu\nu}$ plus a perturbation characterized by the Newtonian Potential whose source $\rho$ is the energy density. 

The only non-zero components of Weyl at this regime are
\begin{eqnarray}
\label{W1} &&{W_{0i}}^{0j}=\partial_i \partial^j\Phi_N\\
\label{W2} &&{W_{ij}}^{kl}=-4 \delta_{[i}^{[k}\partial_{j]} \partial^{l]}\Phi_N
\end{eqnarray}
where $i,j,k,l$ run from $1$ to $3$ and $\Phi_N$ is the Newtonian potential (see \cite{QGPwithoutLSV}). As expected from Weyl's symmetries \cite{Petrov}, expressions (\ref{W1}) and (\ref{W2}) represent the same $3\times3$ real symmetric traceless matrix. The next step is to find $\lambda^{(l)}$ and $q_i^{(l)}$ such that
\begin{equation} \label{lambda y q}
(\partial_i \partial^j\Phi_N) q_j^{(l)}=\lambda^{(l)} q_i^{(l)}.
\end{equation}
In this regime the Weyl eigenforms $\Xi^{(l)}_{\mu\nu}$ and ${\widetilde{\Xi}}^{(l)}_{\mu\nu}$ are related to the $q_i^{(l)}$ via
\begin{equation}
\label{q1} q^{(l)}_i=\Xi^{(l)}_{0i}={\epsilon_i}^{jk}\widetilde{\Xi}^{(l)}_{jk},
\end{equation}
where $\epsilon_{ijk}$ is the totally antisymmetric tensor with $\epsilon_{123}=1$,
and the other components satisfy
\begin{equation}
\label{q2} \Xi^{(l)}_{ij}=\widetilde{\Xi}^{(l)}_{0i}=0.
\end{equation}
Also notice that the conditions (\ref{GXX}) and (\ref{eXX}) will be satisfied if $\delta^{ij}q_i^{(l)}q_i^{(l)}=1$. We have calculated the first post-Newtonian corrections in order to study the effects due to moving sources, but we refer the interested reader to Ref. \cite{QGPwithoutLSV}.

Now we note that the factor $\bar{\psi}\gamma^\mu \gamma^\nu\psi$ also appears in the term $-H_{\mu \nu} \bar{\psi} [\gamma^\mu, \gamma^\nu ] \psi/4$ of the Standard Model Extension (SME) \cite{SME}. This allows us to connect $\mathcal{L}_f$ [given in equation (\ref{L3})] with the SME term by identifying $H_{\mu\nu}$ with
\begin{equation} \label{Hmunu}
H_{\mu\nu}=-2\sum_a \sum_{l=1}^3\sqrt{\lambda^{(l)}}\left\lbrace \xi^{(l)}_a \left( \frac{\sqrt{\lambda^{(l)}}}{M_{Pl}}\right)^r \Xi^{(l)}_{\mu\nu} +{\widetilde{\xi}}^{(l)}_a \left( \frac{\sqrt{\lambda^{(l)}}}{M_{Pl}}\right)^{\widetilde{r}} {\widetilde{\Xi}}^{(l)}_{\mu\nu} \right\rbrace.
\end{equation}
Note however that, in contrast with the SME scheme, the object that plays the role of $H_{\mu \nu}$ is dynamical and depends on the surrounding gravitational sources. Using the formulation of the non-relativistic hamiltonian in the SME \cite{NRH}, the non-relativistic hamiltonian due to the coupling $\mathcal{L}_f$ of a particle with flavor $a$ is
\begin{equation} \label{HNR}
{\cal H}_{NR}= \epsilon^{ijk} \left[\frac{1}{2}\left(\sigma_i + \left(\vec{\sigma}\cdot \frac{\vec{P}}{m} \right) \frac{P_i}{m} \right) H_{jk} + \left(1- \frac{1}{2}\frac{|\vec{P}|^2}{m^2} \right) \frac{P_i}{m} \sigma_j H_{0k} \right],
\end{equation}
where
$\vec{P}$ and $m$ are respectively the momentum and mass of the test particle, the $\sigma_i$ stand for the Pauli matrices, the arrow represent $3$-vector and $\cdot$ is the standard euclidian interior product.

Finally, in order to express the non-relativistic hamiltonian (\ref{HNR}) in a standard $3$-vector notation is useful to define 
\begin{eqnarray}
D_i&\equiv&\frac{1}{2}\epsilon_{ijk}H^{jk}=-2\sum_a\sum_{l=1}^3 \sqrt{\lambda^{(l)}} {\widetilde{\xi}}_a^{(l)} \left( \frac{\sqrt{\lambda^{(l)}}}{M_{Pl}}\right)^{\widetilde{r}}q_i^{(l)},\\
F_i &\equiv& H_{0i}=-2\sum_a \sum_{l=1}^3\sqrt{\lambda^{(l)}}\xi^{(l)}_a \left( \frac{\sqrt{\lambda^{(l)}}}{M_{Pl}}\right)^r q^{(l)}_i,
\end{eqnarray}
where equations (\ref{q1}) and (\ref{q2}) are used. Then, 
\begin{equation}
{\mathcal{H}}_{NR}=\vec{\sigma}\cdot \vec{D}+\left( \vec{\sigma}\cdot \frac{\vec{P}}{M}\right)\left( \vec{D}\cdot \frac{\vec{P}}{M}\right)+\left(1-\frac{1}{2}\frac{P^2}{M^2}\right) \frac{\vec{P}}{M}\cdot \vec{\sigma}\times \vec{F},
\end{equation}
where $\vec{D}$ and $\vec{F}$ are the $3$-vector formed with $D_i$ and $F_i$, respectively, and $\times$ is the standard euclidian exterior product. 

\section{Conclusions and Experimental Outlook}

We have presented a concrete proposal for possible phenomenological manifestations of Quantum Gravity based on the idea that space-time may have a granular structure that respects the Lorentz symmetry. In developing the model, we were forced to introduce the space orientation, through the volume $4$-form, leading us to consider the possibility that  the discrete symmetries, may play a non-trivial role in the fundamental theory of Quantum Gravity. We presented, a non-relativistic hamiltonian that can be used to test the model experimentally. However, before embarking on an experimental search, some important points need to be considered. First, as the Weyl tensor is related to tidal forces, experiments carried out in different locations must be compared with great care. Second, as all the terms of the non-relativistic hamiltonian involve the spin of the particle, the experiments need to involve polarized matter, which must be arranged in a way that the magnetic effects do not hide, or mimic, the effects one is searching for. This seems in principle a very difficult task, however, it should be pointed out that torsion balances with non-magnetic polarized matter have been constructed by the group led by E.G. Adelberger \cite{Adelberger}, opening a window where  this phenomenological model can start to be tested. We must of course look for other situation from where interesting bounds might be obtained.

\end{document}